\documentclass{elsart3p}

\usepackage{graphics, graphicx}
\journal{Physics Letters B}

\begin{document}
\begin{frontmatter}

\title{A first integration of
some knot soliton models}

\author{C. Adam}
\address{Departamento de Fisica de Particulas, Universidad
       de Santiago and Instituto Galego de Fisica de Altas Enerxias
       (IGFAE) E-15782 Santiago de Compostela, Spain}
\ead{adam@fpaxp1.usc.es}

\author{J. S\'{a}nchez-Guill\'{e}n}
\address{Departamento de Fisica de Particulas, Universidad
       de Santiago and Instituto Galego de Fisica de Altas Enerxias
       (IGFAE) E-15782 Santiago de Compostela, Spain}
\ead{joaquin@fpaxp1.usc.es}

\author{A. Wereszczy\'{n}ski}
\address{Institute of Physics,  Jagiellonian University,
       Reymonta 4, 30-059 Krak\'{o}w, Poland}
\ead{wereszczynski@th.if.uj.edu.pl}

\begin{abstract}
Recently it has been shown that there exists a sector within the
Faddeev-Niemi model for which the equations of motion may
be reduced to first order equations. However, no solutions to
that sector have been given.
It is not even known whether this sector contains topologically
nontrivial solutions, at all.
\\
Here, we show that two models with analytically known Hopf
solitons, namely the Nicole and the Aratyn-Ferreira-Zimerman models, possess
sectors which can be integrated to first order partial
differential equations. The main result is that these sectors are
topologically nontrivial. In fact, all analytically known
hopfions belong to them.
\end{abstract}


\end{frontmatter}

\onecolumn
\section{Introduction}
The well-known Faddeev-Niemi model \cite{fn}
is a nonlinear field theory in 3+1 dimensions with the two-sphere S$^2$
as its target space. The maps from one-point compactified three-dimensional
Euclidean space (this compactification is needed for static configurations
to have finite energy) to the target space S$^2$ are classified by a
topological index (the Hopf index), and, therefore, topological solitons
with a knot structure are expected to exist.  \\
The Lagrange density of the Faddeev-Niemi model is
\begin{equation}
\mathcal{L}_{FN}=\mathcal{L}_2 - \lambda \mathcal{L}_4, \label{fn}
\end{equation}
where
\begin{equation}
\mathcal{L}_2 = (\partial_{\mu} \vec{n})^2 = 4
\frac{\partial_{\mu} u \partial^{\mu} \bar{u}}{(1+|u|^2)^2}
\label{l2}
\end{equation}
and
\begin{equation} \label{L4}
\mathcal{L}_4 = [ \vec{n} \cdot (\partial_{\mu} \vec{n} \times
\partial_{\nu} \vec{n})]^2= 4 \frac{(\partial_{\mu} u \partial^{\mu} \bar{u})^2
- (\partial_{\mu} u \partial^{\mu} u)(\partial_{\mu} \bar{u}
\partial^{\mu} \bar{u})}{(1+|u|^2)^4}. \label{l4}
\end{equation}
Moreover, $\lambda$ is a real coupling constant. Here the
topological three component unit vector field $\vec{n}$ is
expressed by a scalar complex $u$ via stereographic projection
\begin{equation}
\vec{n}=\frac{1}{1+|u|^2} \left(u+\bar{u},-i(u-\bar{u}), |u|^2-1
\right). \label{n}
\end{equation}
Due to its nonlinearity and nontrivial topological content the
derivation of knotted solutions in the Faddeev-Niemi model is a
rather difficult problem. Simultaneously with
numerical works, which indicated some knotted configurations as
candidates for the pertinent ground states in sectors with
fixed topological charge \cite{numer1}-\cite{numer5}, some analytical
investigations have been performed. One analytical approach to the FN
model was mainly based on the application of the concept of
integrability to higher dimensions \cite{joaquin1}. In practice it
means that some additional constrains for the complex field are
introduced. As a consequence, one deals with subsectors of the full
theory, which appear to be integrable, in the sense that
infinitely many conserved currents exist \cite{joaquin2}.
Although the constraints, known as integrability conditions,
provide very interesting knotted and linked structures with
arbitrary Hopf index \cite{chris1}, \cite{and1}, these submodels
are still quite complicated and no exact solutions were found up to now.
\\
On the other hand, recently Hirayama and Shi proposed an
alternatively way of solving the FN model. It may be summarize as
follows.
\\
1) The static equation of motion can be written in the form
\begin{equation}
\nabla \cdot \vec{\alpha} + \vec{\beta} \cdot \vec{\alpha}=0,
\label{master eq}
\end{equation}
where the vectors $\vec{\alpha}, \vec{\beta}$ depend on the fields
$u,\bar{u}$ and their derivatives. Notice, that
$\vec{\alpha}$ is not unique. Rather, it can be chosen in
many different ways. But when its particular form is fixed the vector
$\vec{\beta}$ is defined uniquely (For some models there still
remains some freedom for $\vec{\beta}$. This happens, e.g., for the
Aratyn-Ferreira-Zimerman model, see Section 3 for a discussion).
\\
2) Now, the idea is to find vectors $\vec \alpha_s$
which identically satisfy
\begin{equation}
\nabla \cdot \vec{\alpha}_{s} + \vec{\beta} \cdot
\vec{\alpha}_{s}=0, \label{master eq s}
\end{equation}
with a fixed $\vec{\beta}$. On this stage some arbitrary,
'external' complex fields are introduced on which the $\vec \alpha_s$
depend. Solutions are found
using vector identities obeyed by gradients in three dimensional
space. Therefore, one should remember that this procedure works
only in three dimensions.
\\
3) Finally, one has to compare a superposition of the obtained
$\vec{\alpha}_s$'s with the original $\vec{\alpha}$
\begin{equation}
\vec{\alpha}=\sum_{s=1..n} A_i \vec{\alpha}_{s}.
\label{integration eq}
\end{equation}
This equation is a new equation for the complex field which by
construction satisfies the original equation of motion.
\\
Indeed, this approach allows to integrate the equations of motion
to first order partial differential equations.
More precisely, this happens provided that
the simplest nontrivial $\vec{\beta}$ is chosen. By taking
more complicated $\vec{\beta}$, instead, one
arrives at different equations which, in
principle, can be of any order. \\
Let us also emphasize that, although all $\vec{\alpha}$
obeying (\ref{integration eq}) fulfill the original field equations,
the implication in the opposite direction does not have to be true.
Therefore, this procedure defines a submodel of the FN theory. However,
no finite energy solutions to this submodel have been found.
(Unfortunately, the final results in \cite{hirayama} are
not correct due to an error in the derivation,
see \cite{reply-HS} for a detailed discussion.)
It is not even known whether this
submodel allows for any topologically nontrivial solutions at
all.
\\
In our work we apply this framework to two related models where exact Hopf
solitons are known, namely, the Nicole \cite{ni}, \cite{we1},
and the
Aratyn-Ferreira-Zimerman model \cite{afz1}, \cite{afz2}. We show
that they can be integrated to first order equations.
Further, we show that all analytically known hopfions belong, in fact,
to these submodels.
\section{Nicole model}
\subsection{General solutions}
The Nicole model is a scale invariant model built of the kinetic part
$\mathcal{L}_2$ taken to a fractional power
\begin{equation}
\mathcal{L}_{Ni}= (\mathcal{L}_2)^{\frac{3}{2}}. \label{nicole}
\end{equation}
The equations of motion are
\begin{equation}
\nabla \cdot \left[ \frac{(\nabla u \nabla
\bar{u})^{\frac{1}{2}}}{(1+|u|^2)^3} \nabla \bar{u} \right] +
\frac{2\bar{u}}{(1+|u|^2)^4} (\nabla u \nabla
\bar{u})^{\frac{3}{2}}=0 \label{static eom}
\end{equation}
and its complex conjugate. Now, let us introduce a function $g$
which is assumed to depend on $|u|^2$ only. In principle one may
consider a more general case when $g$ is a function of $u$ and
$\bar{u}$ independently, or allow for dependence on derivatives.
This last possibility leads to higher than first order partial
differential equation for the complex field. In this paper we restrict
ourselves to the first case.
\\
Now, the last expression can be rewritten as
\begin{equation}
\nabla \cdot \left[ g (\nabla u \nabla \bar{u})^{\frac{1}{2}}
\nabla \bar{u} \right] + g (\nabla u \nabla \bar{u})^{\frac{1}{2}}
\nabla \bar{u} \cdot \left[ \nabla u \left( -\frac{g'\bar{u}}{g} -
\frac{\bar{u}}{1+|u|^2}\right) + \nabla \bar{u} \left(
-\frac{g'u}{g} - \frac{3u}{1+|u|^2} \right) \right]=0.
\label{static eom mod2}
\end{equation}
Here the prime denotes differentiation with respect to $|u|^2$. In
other words, we expressed the equation of motion in the required form
$$\nabla \cdot \vec{\alpha} + \vec{\beta} \cdot \vec{\alpha}=0 $$
with the following vectors
\begin{equation}
\vec{\alpha}= g (\nabla u \nabla \bar{u})^{\frac{1}{2}} \nabla
\bar{u}, \label{alpha def}
\end{equation}
\begin{equation}
\vec{\beta}=-\left(\frac{g'\bar{u}}{g} +
\frac{\bar{u}}{1+|u|^2}\right) \nabla u - \left(\frac{g'u}{g} +
\frac{3u}{1+|u|^2} \right) \nabla \bar{u}. \label{beta def}
\end{equation}
After the polar decomposition
\begin{equation}
u=Re^{i\Phi} \label{polar}
\end{equation}
we get
\begin{equation}
\vec{\beta}=-2\left(\frac{g'}{g}+\frac{2}{1+R^2} \right) R \nabla
R + \frac{2iR^2}{1+R^2} \nabla \Phi. \label{beta polar}
\end{equation}
Here, the prime stands for differentiation with respect to $R^2$.
Now, we solve equation (\ref{master eq}) by introducing 'external' complex
fields $\mu$ and $\rho$, analogous to those of Ref.
\cite{hirayama}. There are two solutions
\begin{equation}
\vec{\alpha_1}=\nabla \Phi \times \nabla \mu + i \left(
\frac{g'}{g}+\frac{2}{1+R^2} \right) \frac{1+R^2}{R} \nabla R
\times \nabla \mu \label{alpha1}
\end{equation}
\begin{equation}
\vec{\alpha_2}= K(\Phi) \; g(1+R^2)^2 \nabla \Phi \times \nabla
\rho +  \tilde{K}(R) \; e^{-2i \frac{R^2}{1+R^2} \Phi} \nabla R
\times \nabla \rho. \label{alpha2}
\end{equation}
$K$ and $\tilde{K}$ are arbitrary complex functions of $\Phi$ and $R$,
respectively. There is also a third solution perpendicular to the
gradients of $\Phi$ and $R$ which does not depend on those
external complex functions
\begin{equation}
\vec{\alpha_3}=G(R,\Phi) \nabla \Phi \times \nabla R.
\label{alpha3}
\end{equation}
Interestingly,  $\vec{\alpha}_1$ and $\vec{\alpha}_3$ even solve
the stronger equation
\begin{equation}
\nabla \cdot \vec{\alpha}=0, \;\;\; \vec{\beta} \cdot
\vec{\alpha}=0. \label{submodel}
\end{equation}
Moreover, one can observe that the second solution can be
trivially generalized to
\begin{equation}
\vec{\alpha_2}= K(\Phi, \rho) \; g(1+R^2)^2 \nabla \Phi \times
\nabla \rho +  \tilde{K}(R, \rho) \; e^{-2i \frac{R^2}{1+R^2}
\Phi} \nabla R \times \nabla \rho. \label{alpha_2gen}
\end{equation}
To summarize this subsection, we have derived the following first
order partial differential equation whose solutions identically
solve the original second order equations of motion
\begin{eqnarray}
 g \sqrt{(\nabla R)^2+R^2 (\nabla \Phi)^2} [\nabla R - iR
\nabla \Phi] e^{-i\Phi}= \\ A_1 \left[ \nabla \Phi \times \nabla
\mu + i \left( \frac{g'}{g}+\frac{2}{1+R^2} \right)
\frac{1+R^2}{R} \nabla R \times \nabla \mu\right]+ \label{mod foeq
1}
\\
 A_2 \left[K(\Phi, \rho) \; g(1+R^2)^2 \nabla \Phi
\times \nabla \rho +  \tilde{K}(R, \rho) \; e^{-2i
\frac{R^2}{1+R^2} \Phi} \nabla R \times \nabla \rho \right] +
\label{mod foeq 2}
\\
 A_3 \; G(R,\Phi) \nabla \Phi \times \nabla R,
\label{mod foeq 3}
\end{eqnarray}
where $A_1,A_2,A_3$ are complex constants.
\\
The standard Hirayama and Shi type solution can be rederived if we assume
$g=(1+|u|^2)^{-2}$ and $K=\tilde{K}=1, G=0$. Then
\begin{equation}
\vec{\alpha_1}=\nabla \Phi \times \nabla \mu \label{alpha1 hira}
\end{equation}
\begin{equation}
\vec{\alpha_2}=\nabla \Phi \times \nabla \rho + e^{-2i
\frac{R^2}{1+R^2} \Phi} \nabla R \times \nabla \rho. \label{alpha2
hira}
\end{equation}
Thus the pertinent vector is
\begin{equation}
\vec{\alpha}_H=\vec{\alpha}_1(\mu) +
\vec{\alpha}_2(\rho)-\vec{\alpha}_1(\rho)= \nabla \Phi \times
\nabla \mu + e^{\frac{-2iR^2}{1+R^2}\Phi} \nabla R \times \nabla
\rho. \label{alpha hira}
\end{equation}
\subsection{$Q=1$ hopfion}
In this part of the paper we want to discuss the problem whether
the well-known soliton solution with topological charge one does
or does not belong to the solutions of the first order equation
(\ref{mod foeq 1})-(\ref{mod foeq 3}).
As we will see it is sufficient to consider the simplest case
(\ref{alpha hira}). \\
For the discussion of the unit charge Hopf soliton it is useful to
introduce toroidal coordinates,
\begin{eqnarray}
x &=&  q^{-1} \sinh \eta \cos \varphi \;\;, \;\;
y =  q^{-1} \sinh \eta \sin \varphi   \nonumber \\
z &=&  q^{-1} \sin \xi \quad ;  \qquad  q = \cosh \eta - \cos \xi .
\label{tordefs}
\end{eqnarray}
Then the Hopf soliton takes the form
\begin{equation}
u=\frac{1}{\sinh \eta} e^{i(\varphi+\xi)}. \label{soliton}
\end{equation}
Thus
\begin{equation}
R=\frac{1}{\sinh \eta}, \;\;\; \Phi=\xi + \varphi. \label{soliton1}
\end{equation}
The condition
$\vec{\alpha}=\vec{\alpha}_H $ calculated for the hopfion
is equivalent to the following three first
order partial but linear differential equations
\begin{equation}
-\sqrt{2} \frac{1}{\cosh^2 \eta} e^{-i(\varphi+\xi)}=\frac{1}{\sinh
\eta} (\mu_{\varphi}-\mu_{\xi}), \label{q1H eq 1}
\end{equation}
\begin{equation}
-i\sqrt{2} \frac{\sinh \eta}{\cosh^3 \eta} e^{-i(\varphi+\xi)}=
\frac{1}{\sinh \eta} \mu_{\eta} + \frac{\cosh \eta}{\sinh^3 \eta}
e^{\frac{-2i}{\cosh^2 \eta}(\xi+\varphi)} \rho_{\varphi}, \label{q1H eq
2}
\end{equation}
\begin{equation}
-i\sqrt{2} \frac{1}{\cosh^3 \eta} e^{-i(\varphi+\xi)}= - \mu_{\eta} -
\frac{\cosh \eta}{\sinh^2 \eta} e^{\frac{-2i}{\cosh^2
\eta}(\xi+\varphi)} \rho_{\xi}. \label{q1H eq 3}
\end{equation}
The first equation (\ref{q1H eq 1}) has the following solution
\begin{equation}
\mu= \tilde{\mu} (\eta, \varphi+\xi) -\sqrt{2} \frac{\sinh
\eta}{\cosh^2 \eta} e^{-i(\varphi+\xi)} \varphi, \label{q1H sol mu}
\end{equation}
where $\tilde{\mu}$ is an arbitrary function of $\eta$ and
$\varphi+\xi$. Moreover, if we multiply (\ref{q1H eq 2}) by $\sinh
\eta$ and add it to (\ref{q1H eq 3}), then we obtain an equation for
$\rho$. Namely,
\begin{equation}
-i\sqrt{2} \frac{\sinh^2 \eta}{\cosh^2 \eta} e^{-i\frac{\sinh^2
\eta-1}{\cosh^2 \eta} (\xi+\varphi)} = \rho_{\varphi}-\rho_{\xi}.
\label{q1H eq 4}
\end{equation}
The corresponding solution reads
\begin{equation}
\rho = \tilde{\rho}(\eta, \xi+\varphi) -i\sqrt{2} \frac{\sinh^2
\eta}{\cosh^2 \eta} e^{-i\frac{\sinh^2 \eta-1}{\cosh^2 \eta}
(\xi+\varphi)} \varphi. \label{q1H sol rho}
\end{equation}
The last remaining step is to insert these solutions, e.g., into the
third equation (\ref{q1H eq 3}). One finds that
\begin{eqnarray}
-i\sqrt{2} \frac{1}{\cosh^3 \eta} e^{-i(\xi+\varphi)}
+\tilde{\mu}_{\eta} + \frac{\cosh \eta}{\sinh^2 \eta}
e^{-i\frac{\sinh^2 \eta-1}{\cosh^2 \eta} (\xi+\varphi)}
\tilde{\rho}_{\xi}=\sqrt{2}e^{-i(\xi+\varphi)} \varphi \times \\ \left[
\left(\frac{\sinh \eta}{\cosh^2 \eta} \right)' + \frac{\cosh
\eta}{\sinh^2 \eta } \frac{\sinh^2 \eta}{\cosh^2 \eta}
\frac{\sinh^2-1}{\cosh^2 \eta} \right]. \label{q1H consist}
\end{eqnarray}
However, the right hand side of this equation is equal to zero, because the
expression in brackets vanishes identically. In other words,
the unit hopfion does belong to the Hirayama and Shi subsector, i.e., the
hopfion is a solution of the first order PDE with $\mu$, $\rho$
given by (\ref{q1H sol mu}) and (\ref{q1H sol rho}) respectively,
where $\tilde{\mu}$ and $\tilde{\rho}$ obey
\begin{equation}
-i\sqrt{2} \frac{1}{\cosh^3 \eta} e^{-i(\xi+\varphi)}
+\tilde{\mu}_{\eta} + \frac{\cosh \eta}{\sinh^2 \eta}
e^{-i\frac{\sinh^2 \eta-1}{\cosh^2 \eta} (\xi+\varphi)}
\tilde{\rho}_{\xi}=0. \label{nicole sol cond}
\end{equation}
\section{AFZ model}
Let us now consider another scale invariant model, namely the so-called
Aratyn-Ferreira-Zimerman model presented in the introduction.
\subsection{General solutions}
With $\mathcal{L}_4$ given by eq. (\ref{L4}), the Lagrangian reads
\begin{equation}
\mathcal{L}_{AFZ}=-(\mathcal{L}_4)^{\frac{3}{4}}. \label{afz}
\end{equation}
The equations of motion are
\begin{equation}
\nabla \left[ \frac{(\vec{K} \nabla u)^{-\frac{1}{4}}
}{(1+|u|^2)^3} \vec{K} \right] + (\vec{K} \nabla u)^{\frac{3}{4}}
\frac{2\bar{u}}{(1+|u|^2)^4}=0 \label{afz eom 1}
\end{equation}
and its complex conjugate. Here
\begin{equation}
\vec{K}=(\nabla \bar{u})^2 \nabla u - (\nabla u \nabla \bar{u})
\nabla \bar{u}. \label{k def}
\end{equation}
Introducing an arbitrary function of the modulus, $g=g(|u|^2)$, the
field equation may be rewritten as follows
\begin{equation}
\nabla [g (\vec{K} \nabla u)^{-\frac{1}{4}} \vec{K}] + g (\vec{K}
\nabla u)^{-\frac{1}{4}}
\bar{u}\left(-\frac{g'}{g}-\frac{1}{1+|u^2} \right) \vec{K} \cdot
\nabla u=0. \label{afz eom 2}
\end{equation}
Thus, again we arrive at
\begin{equation}
\nabla \cdot \vec{\alpha} + \vec{\beta} \cdot \vec{\alpha}=0,
\label{afz eom gen}
\end{equation}
where
\begin{equation}
\vec{\alpha}=g (\vec{K} \nabla u)^{-\frac{1}{4}} \vec{K}
\label{afz alpha def}
\end{equation}
and
\begin{equation}
\vec{\beta}= - \bar{u}\left(\frac{g'}{g}+\frac{1}{1+|u^2} \right)
\nabla u. \label{afz beta def}
\end{equation}
Observe that $\vec{K}$ satisfies the identity
\begin{equation}
\vec{K} \cdot \nabla \bar{u} \equiv 0. \label{k identity}
\end{equation}
Thus, one can always include in $\vec{\beta}$ a part which is
proportional to the gradient of $u$. That is, the most general
$\vec{\beta}$ reads
\begin{equation}
\vec{\beta}= - \left(\frac{g'}{g}+\frac{1}{1+|u|^2} \right)
[\bar{u} \nabla u + h \nabla \bar{u}], \label{afz beta gen}
\end{equation}
where $h$ is any function depending on arbitrary variables.
However, for simplicity, in the subsequent analysis we put
$h=-1$. Therefore, after the polar decomposition
\begin{equation}
\vec{\beta}= - 2iR^2 \left(\frac{g'}{g}+\frac{1}{1+R^2} \right)
\nabla \Phi. \label{afz beta}
\end{equation}
Similarly as for the Nicole model, we solve equation (\ref{afz eom
gen}) for fixed $\vec{\beta}$. There are three solutions
\begin{equation}
\vec{\alpha}_1=\nabla \Phi \times \nabla \mu, \label{afz alpha1}
\end{equation}
\begin{equation}
\vec{\alpha}_2=e^{2i R^2 \left(\frac{g'}{g}+\frac{1}{1+R^2}
\right)\Phi} \nabla R \times \nabla \rho, \label{afz alpha2}
\end{equation}
and
\begin{equation}
\vec{\alpha}_3=G(\Phi,R)\nabla \Phi \times \nabla R. \label{afz
alpha3}
\end{equation}
Choosing a particular form for the $g$ function
\begin{equation}
g=\frac{1}{(1+|u|^2)^3}
\end{equation}
we find the Hirayama and Shi type solution
\begin{equation}
\vec{\alpha}_H=\nabla \Phi \times \nabla \mu + e^{-4i
\frac{R^2}{1+R^2} \Phi} \nabla R \times \nabla \rho . \label{afz
hira}
\end{equation}
In the next subsection we prove that the Hopf solitons can be
indeed expressed in this form.
\subsection{Hopfions}
It is widely known that the AFZ model possesses infinitely many
analytically known finite energy toroidal
Hopf solitons which, in toroidal coordinates, read
\begin{equation}
u=f e^{i(m\xi + n\varphi)}. \label{afz hopfion}
\end{equation}
Here
\begin{equation}
f^2= \frac{\cosh \eta - \sqrt{n^2/m^2 + \sinh^2
\eta}}{\sqrt{1+m^2/n^2 \sinh^2 \eta}-\cosh \eta}, \label{afz f}
\end{equation}
whereas $m,n$ are integer constants. The corresponding topological
charge is $Q=-nm$. Hence, in our parametrization
\begin{equation}
R=f, \;\;\; \Phi=m\xi + n \varphi. \label{afz hopf R Phi}
\end{equation}
Now, we establish that
\begin{equation}
\vec{\alpha}=\vec{\alpha}_H \label{afz cond}
\end{equation}
or in other words, that these hopfions belong to Hirayama and Shi
submodel. \\
The last formula leads to three first order equations. Namely,
\begin{equation}
-\sqrt{2} (f'f)^{\frac{1}{2}} \frac{f}{(1+f^2)^3} \left(
m^2+\frac{n^2}{\sinh^2 \eta} \right)^{\frac{3}{4}}
e^{-i(m\xi+n\varphi)}= \frac{1}{\sinh \eta} (m\mu_{\varphi}-n\mu_{\xi})
\label{afz hopf eq1}
\end{equation}
\begin{equation}
i\sqrt{2}m (f'f)^{\frac{1}{2}} \frac{f'}{(1+f^2)^3} \left(
m^2+\frac{n^2}{\sinh^2 \eta} \right)^{-\frac{1}{4}}
e^{-i(m\xi+n\varphi)}=
 \frac{n}{\sinh \eta} \mu_{\eta} - \frac{f'}{\sinh
\eta} \rho_{\varphi} e^{-4i \frac{f^2}{1+f^2} (m\xi+n\varphi)},
\label{afz hopf eq2}
\end{equation}
\begin{equation}
 i\sqrt{2}n (f'f)^{\frac{1}{2}}
\frac{f'}{(1+f^2)^3 \sinh \eta } \left( m^2+\frac{n^2}{\sinh^2
\eta} \right)^{-\frac{1}{4}} e^{-i(m\xi+n\varphi)}=
 -m \mu_{\eta} +f'\rho_{\xi} e^{-4i
\frac{f^2}{1+f^2} (m\xi+n\varphi)}. \label{afz hopf eq3}
\end{equation}
The first equation may be integrated and gives
\begin{equation}
\mu=\tilde{\mu}(\eta, n\varphi+m\xi) - \sqrt{2} (f'f)^{\frac{1}{2}}
\frac{f}{(1+f^2)^3} \left( m^2+\frac{n^2}{\sinh^2 \eta}
\right)^{\frac{3}{4}} e^{-i(m\xi+n\varphi)} \frac{\varphi}{m}.
\label{afz mu sol}
\end{equation}
Moreover, we multiply (\ref{afz hopf eq2}) by $m \cosh \eta$ and
add it to (\ref{afz hopf eq3}). Then we get
\begin{equation}
 -i\sqrt{2} (f'f)^{\frac{1}{2}}
\frac{f'}{(1+f^2)^3} \left( m^2+\frac{n^2}{\sinh^2 \eta}
\right)^{\frac{3}{4}} \sinh \eta e^{-i(m\xi+n\varphi)}= (m\rho_{\varphi}
- n \rho_{\xi}) e^{-4i \frac{f^2}{1+f^2} (m\xi+n\varphi)} , \label{afz
hopf eq4}
\end{equation}
which has the following solution
\begin{equation}
\rho= \tilde{\rho}(\eta,m\xi+n\varphi)-i\sqrt{2} (f'f)^{\frac{1}{2}}
\frac{f'\sinh \eta}{(1+f^2)^3} \left( m^2+\frac{n^2}{\sinh^2 \eta}
\right)^{\frac{3}{4}} e^{-i(1-\frac{4f^2}{1+f^2})
(m\xi+n\varphi)}\frac{\varphi}{m}. \label{afz rho sol}
\end{equation}
Finally, inserting the obtained expressions into the third equation
(\ref{afz hopf eq3}) we arrive at
\begin{eqnarray}
\frac{i\sqrt{2}n (f'f)^{\frac{1}{2}}f'}{(1+f^2)^3 \sinh \eta}
\left( m^2+\frac{n^2}{\sinh^2 \eta} \right)^{-\frac{1}{4}}
e^{-i(m\xi+n\varphi)} +m\tilde{\mu}_{\eta} - f'
\tilde{\rho}_{\xi}e^{-4i \frac{f^2}{1+f^2} (m\xi+n\varphi)}=
 \sqrt{2}e^{-i(m\xi+n\varphi)} \varphi \times \\ \left[ \left(
\frac{(f'f)^{\frac{1}{2}}f\sinh \eta}{(1+f^2)^3} \left(
m^2+\frac{n^2}{\sinh^2 \eta} \right)^{\frac{3}{4}}\right)'_{\eta}
\right.
 \left. - \frac{(f'f)^{\frac{1}{2}}f'\sinh
\eta}{(1+f^2)^3} \left( m^2+\frac{n^2}{\sinh^2 \eta}
\right)^{\frac{3}{4}} \left( 1-\frac{4f^2}{1+f^2}\right)\right]
\end{eqnarray}
However, the right hand side of this last formula is zero if the
expression for $f$ is taken into account.
To conclude, we have demonstrated that all
hopfion solutions are solutions of the first order PDE
defined by $\vec{\alpha}_H$.
\section{Conclusions}
The main achievement of this paper is the demonstration that all
known exact Hopf solitons in the Nicole as well as the AFZ models
may be expressed by solutions of the first order equations of
Hirayama and Shi with $\mu$ and $\rho$ defined in the previous
sections. As a consequence, this method of the reduction of the
equations of motion appears to be applicable to models with
knotted solitons, because topologically nontrivial solutions of
the theory are taken into account by the Ansatz. Therefore, this
framework probably will incorporate knotted configurations for the
Faddeev-Niemi model, as well, and may be useful in the search for
solutions or in further analytical investigations of that model.

Let us further remark that the method
can be easily generalized to models with a
potential.
For example, if one adds a potential term $24V(uu^*)$ to the Nicole model,
then this just modifies $\vec \beta$ to
$$
\vec{\beta}= \vec{\beta}_{old} - (1+|u|^2)^3 V' \frac{u^* \nabla u}{(\nabla u
\nabla u^*)^(3/2)}.
$$
This modification is quite simple, but since for realistic applications of the
FN model a potential term is needed,
this makes the
procedure all the more interesting. Now, $\vec \beta$ depends on derivatives,
but in the vectorial sense it is still proportional to
$\nabla u$ and $\nabla u^*$.

Further generalizations are provided by allowing for
more complicated $g$. In principle, one may consider a case
where $g$ is a function not only of the modulus but depends on $u,\bar{u}$
in an arbitrary way. This possibility might lead to configurations
without toroidal symmetry which are quite typical for the FN model
like, for instance,  trefoil knots. Moreover, one can also allow for a
dependence on (higher) derivatives. Then, after constructing
solutions of Eq. (\ref{master eq}) and comparing them with
the primary $\vec{\alpha}$, one gets a more complicated set of
partial differential equations. If $g$ depends on first
derivatives, the resulting PDE is of the second order. In general,
we are able to find PDE of any order whose solutions identically
solve the original theory. This might allow for the construction of
a hierarchy of submodels.

\ack C.A. and J.S.-G. thank MCyT (Spain) and FEDER
(FPA2005-01963), and support from Xunta de Galicia (grant
PGIDIT06PXIB296182PR and Conselleria de Educacion). A.W. is
partially supported from Jagiellonian University (grant WRBW
41/07). Further, C.A. acknowledges support from the Austrian START
award project FWF-Y-137-TEC and from the FWF project P161 05 NO 5
of N.J. Mauser.


\thebibliography{45}

\bibitem{fn} Faddeev L and Niemi N 1999 Phys. Rev. Lett. {\bf 82} 1624
\bibitem{numer1} Gladikowski J and Hellmund M 1997 Phys. Rev. D {\bf
56} 5194
\bibitem{numer2} Battye R A and Sutcliffe P M 1998 Phys. Rev. Lett. {\bf
81} 4798
\bibitem{numer2a} Battye R A and Sutcliffe P M 1999 Proc.Roy.Soc.Lond. A
{\bf 455}
4305
\bibitem{numer3} Hietarinta J and Salo P 1999 Phys. Lett. B {\bf 451} 60
\bibitem{numer4} Hietarinta J and Salo P 2000 Phys. Rev. D {\bf 62} 81701
\bibitem{numer5} Sutcliffe P M 2007 arXiv:0705.1468
\bibitem{joaquin1} Alvarez O, Ferreira L A and S\'{a}nchez-Guill\'{e}n
J 1998 Nucl. Phys. B {\bf 529} 689
\bibitem{joaquin2} S\'{a}nchez-Guill\'{e}n J 2002 Phys. Lett. B {\bf 548} 252,
Erratum-ibid. B {\bf 550} 220
\bibitem{chris1}  Adam C 2004 J. Math. Phys. {\bf 45} 4017
\bibitem{and1} Wereszczy\'{n}ski A 2005 Eur. Phys. J. C {\bf 42} 461
\bibitem{hirayama} Hirayama M and Shi C-G 2007 Phys. Lett. B {\bf
652} 384
\bibitem{reply-HS}
 Adam C, S\'{a}nchez-Guill\'{e}n J, and
Wereszczy\'{n}ski A 2007 arXiv:0710.4088
\bibitem{ni} Nicole D A 1978 J. Phys. G {\bf 4} 1363
\bibitem{we1}  Adam C, S\'{a}nchez-Guill\'{e}n J,
V\'azquez R, and
Wereszczy\'{n}ski A 2006 J. Math. Phys. {\bf 47} 052302
\bibitem{afz1} Aratyn H, Ferreira L A and Zimerman A H 1999 Phys.
Lett. B {\bf 456} 162
\bibitem{afz2} Aratyn H, Ferreira L A and Zimerman A H 1999
Phys. Rev. Lett. {\bf 83} 1723
\end{document}